# MULTI-BAND IMAGES
# OF THE BARRED GALAXY
# NGC 1097

A.C. Quillen,[1] Jay. A. Frogel,[2] L. E. Kuchinski,[3] D. M. Terndrup,[3,4]

The Ohio State University, Department of Astronomy
174 West 18th Avenue, Columbus, OH 43210

January 4, 1995

[1] E-mail: quillen@payne.mps.ohio-state.edu

[2] Visiting Associate of the Observatories, Carnegie Institution of Washington

[3] Visiting Astronomer, Cerro Tololo International Observatory

[4] Presidential Young Investigator

## ABSTRACT

We present $B, V, R, J, H, K$ broad–band images of the barred galaxy NGC 1097. The optical and infrared colors maps trace the location of the major dust features. The dust lanes are relatively free of star formation and have low opacity. The depth of the dust lanes decreases as a function of wavelength so that they are deepest at $B$ and shallower at $H$ and $K$ where they are difficult to detect. We find that the dust lane on the north-west side of the galaxy has colors consistent with those of a dust screen with opacity derived from the galactic extinction law. Thus, it must be located on the near side of the galaxy. The colors in the dust lane on the south-east side of the galaxy, on the other hand, suggest that there are stars both in front of and behind the dust, consistent with this dust lane being located on the far side of the galaxy.

From the $K$ images, we estimate that the total stellar mass of the star forming spiral ring at a radius of $\sim$ 1kpc from the nucleus is of the same order as the molecular gas mass in the ring. The infrared images show a short bar inside the inner spiral ring. In the principal plane of the galaxy, the short bar is not perpendicular to the prominent outer bar. This suggests that there are torques between the inner bar and the spiral ring and that this inner bar may have a pattern speed different from the outer bar.

Near the inner spiral ring (at $\sim 17'' = 1.4$kpc from the nucleus), the dust lane becomes double peaked with peak to valley extinction ratio similar to that observed in radio continuum by Ondrechen and van der Hulst (1983). This suggests that if the cold diffuse component of the interstellar medium (traced by the dust) moves with the cosmic ray electron component then the magnetic field pressure is not a significant force in the shocks. Alternatively if the two phases do not move together a more detailed comparison may show differences in the shock shapes.

*Subject headings:* galaxies: individual (NGC 1097) – dust, infrared images

## 1. INTRODUCTION

The nearby southern barred galaxy NGC 1097 is noted for its large scale radio quiet optical jets extending out to $\sim$ 90 kpc (Wolstencroft and Zealy 1975, Lorre 1978, Wolstencroft, Tully, & Perley 1984). Its nucleus has a compact radio source and an emission-line spectrum which classifies it as a LINER or a weak Seyfert (Talent 1982, Keel 1983, Phillips et al. 1984, Véron-Cetty & Véron 1986) with a recently discovered obscured broad-line region (Storchi-Bergman, Baldwin & Wilson 1993). NGC 1097 is also noted for a nuclear ring at a radius of $\sim 10'' = 800$pc which is the site of enhanced star formation (Hummel, van der Hulst & Keel 1987) and supernova 1992bd (Chris Smith, private communication; International Astronomical Union, Circular #5638, 1992). The ring has a molecular gas mass of $\sim 1.3 \times 10^9 M_\odot$ $H_2$ (Gerin, Nakai & Combes 1988) and is resolved into tightly wound spiral arms in the near IR (Forbes et al. 1992). The smooth, thin dust lanes running along the leading side of the bar are coincident with radio continuum emission (Ondrechen & van der Hulst 1983). HI kinematics measured in the bar show large scale non-circular motions (Ondrechen, van der Hulst & Hummel 1989).

In this paper we report the results of an optical and near-infrared study of NGC 1097. In §2 we present images of NGC 1097 at $B(0.44\mu m)$, $V(0.55\mu m)$, $R(0.71\mu m)$, $J(1.25\mu m)$, $H(1.65\mu m)$, and $K(2.2\mu m)$. As discussed in §3, the optical and infrared color maps show that the dust lanes are smooth, and are nearly free of young stars. Thus, the broad base line of our observations provides an opportunity to investigate the optical properties of the dust as a function of wavelength without the added confusion of young star clusters. The structure of the dust lanes themselves should trace changes in the diffuse cool component of the interstellar medium (ISM) which is expected to be engaged in a shock-like response to the barred galactic potential. Our data are a preliminary part of a survey of 200 to 300 galaxies that will produce a library of photometrically calibrated images of late-type galaxies from 0.4 to 2.2$\mu$m. For $H_0 = 75$ km s$^{-1}$ Mpc$^{-1}$, NGC 1097 is at a distance of 17 Mpc at which scale $1''$ corresponds to $\sim 80$ pc.

## 2. OBSERVATIONS

### 2.1 Near Infrared Images J, H, & K

The infrared images were obtained on the 1.0m Swope Telescope at Las Campanas Observatory 1992 Oct. 31, 1992 Nov. 4, 7, and 1993 Oct. 3. The infrared camera (Persson et al. 1992) covered a field of 3.9×3.9 arcminutes, with a spatial scale of 0.92 arcsec/pixel. This data set consists primarily of 20-40 minute on-source integrations at 5 positions of the galaxy. One position is centered on the nucleus while the other four have the nuclear region in the corner of the array so that regions of overlap were substantial. Individual images were taken with an exposure time of 15,30,60s at $J$, $H$ and $K$ respectively. Total exposure times for each of the five positions at $J$ were $\sim$20 minutes. At $K$ the total exposure times were $\sim$20 minutes for the central image and the north-west position and $\sim$40 minutes at the other 3 positions. An $H$ image at the north-west position was observed on 1993 Oct 3 with a total integration time of 18 minutes. A log of the observations is given in Table 1.

The sky was observed for a total integration time of about half of the total on source integration time. Flat fields were constructed from median filtered sky frames. The individual images at each position were aligned based on centroids of stars in the field and the position of the nucleus to the nearest pixel and combined after a slight non-linearity correction, flat fielding and sky subtraction. Raw data images are flat to a few % even without flat fielding (Persson et al. 1992). A planar surface was removed from the resulting images to correct for problems in sky subtraction probably due to scattered light.

The images were combined to make mosaics in $J$ and $K$ which completely cover the bar of NGC 1097 and the neighboring dwarf elliptical galaxy NGC 1097A. The mosaics were constructed by shifting the images observed at the 5 positions according to position of the nucleus. A relative multiplication factor and the base sky level for each image was adjusted to minimize differences in the images in the overlap regions. The multiplication factors varied by less than 4% (from the mean multiplication factor) for $K$ images and less than 2% for the $J$ images. Stars in the two mosaics have $\sim 2.6''$ FWHM. The single $H$ image has slightly worse resolution with stars of $\sim 2.9''$ FWHM.

Short 2 minute exposures in all 3 bands on 1993 Oct 3 were observed under excellent photometric conditions and were used to calibrate the long exposure IR images. Comparison of the two sets of images showed no variation in the luminosity of the nucleus and that supernova 1992bd contributed $\ll 1\%$ to the luminosity of the nuclear ring. Our calibration was therefore unaffected by variability in these sources. The photometric images were calibrated using standard stars from Elias et al. (1982) to convert to the CTIO/CIT system. We compared the resulting images to aperture photometry by Glass & Moorwood (1985) with 7 aperture diameters and find that our $J$, $H$ and $K$ magnitudes agree within 0.04 magnitudes with theirs. In our images the noise per $0.92''$ pixel has a standard deviation that corresponds to 20.2, 20.8, 21.6 mag arcsec$^{-2}$ in the $J$ mosaic, $H$ image, and $K$ mosaic respectively. Near infrared images are displayed in Figures 1 − 4.

## 2.2 Optical Images $B$, $V$, & $R$

The $B$, $V$, and $R$ images were observed under photometric conditions on UT 1993 Dec. 12 on the 0.9m telescope at CTIO, using the $1024 \times 1024$ pixel CCD with a spatial scale of $0.40''$/pixel. The images were taken with an exposure time of 900 seconds in $R$, 900 seconds in $V$ and 1200 seconds in $B$. Flat fields were constructed from means of dome flats. Images were flat fielded after dark subtraction and a linear overscan bias subtraction. Cosmic rays events were removed using interactive median filtering.

Calibration was done by observing stars in 2 E-regions and converting to the $UBVRI$ system using the photometry of these regions presented in Graham (1982). From the scatter of magnitudes of the standard stars we estimate our photometric accuracy to be better than 0.06 magnitudes. We compared magnitudes measured in synthetic round apertures centered on the nucleus in our $B$ and $V$ images to those listed listed in Longo & deVaucouleurs (1983). We find that our $B$ and $V$ zero points agree within 0.15 mag to all magnitudes listed in Longo & deVaucouleurs (1983). Stars in the original images have FWHM $\approx 1.7''$. Optical images that are resampled to the same pixel scale, orientation and resolution (FWHM=$2.9''$) as the infrared images are displayed in Figures 1 − 4. In these images the noise per pixel has a standard deviation that corresponds to 24.0 mag arcsec$^{-2}$ in the three optical bands.

## 2.3 Color maps

Registration for the color maps was done by centroiding on stars in the field. To tranform all images to the same pixel scale, orientation and resolution, we used linear transformations which included a rotation and a scaling term. Images were transformed to the scale of the $0.92''$/pixel infrared images. We estimate that our registration is good to within $0.3''$ based upon the variance of centroids in different images of the field stars of the registered images. Prior to division the images were smoothed so that stars had the same FWHM as the $H$ image. Color maps are displayed in Figures 2, 3 and 4.

## 3. RESULTS

### 3.1 *Overall Morphology*

Figure 1 displays our infrared and optical images smoothed to the same spatial resolution of NGC 1097 and its dwarf elliptical companion, NGC 1097A, to the north west of NGC 1097. The infrared images appear to be much smoother than the optical images. In particular, except for the central ring and nucleus, the infrared images contain no sharp features. The galaxy is approximately at an inclination of 45° (Ondrechen, van der Hulst, & Hummel 1989) so that at any position on the sky we see stars at a range of distances from the galactic plane of NGC 1097. Components that have a large vertical scale height will appear smoother than those with a small vertical scale height. Dust generally has a vertical scale height that is smaller than the stellar scale height so that features caused by absorption from dust in the optical images are expected to have fine structure in our images, as we observe. Groupings of young luminous stars should have a small scale height and form sharp features in the optical images. An older stellar population, on the other hand, should have a larger scale height, appear more evenly distributed, and dominate the near infrared images (Frogel 1988).

### 3.2 *NGC 1097A*

Figure 2 shows enlargements and color maps of NGC 1097A, the small dwarf elliptical galaxy to the north west of NGC 1097. Although the galaxy shows a strong radial surface brightness gradient, the colors in this dwarf elliptical galaxy are uniform. This implies that that there is little dust in this galaxy. The near infrared colors are bluer than the colors in the bar of NGC 1097 outside the dust lanes (in Table 2 compare the column labeled 'NGC 1097A' with the column labeled 'bar'). NGC 1097A has $B - H \approx 3.5$, $J - K \approx 0.77$ and a total $H$ magnitude of 10.7, which at our assumed distance implies an absolute magnitude of −20.3 mags. These colors and the absolute magnitude are are similar to those of the Virgo dwarf elliptical galaxies studied by Bothun et al. (1986). The outer isophotes of NGC 1097A appear to be slightly peanut shaped which may be caused by tidal stripping. Note the arm-like feature apparent in both the optical and infrared images (see Figure 1), which extends from the north west side of NGC 1097 and curves northwards up towards NGC 1097A. We suspect that this feature is a tidal tail of NGC 1097A.

### 3.3 *The nuclear spiral-ring*

Figure 3 shows that a ring of enhanced (red) colors are observed in the near infrared color maps coincident with the spiral pattern seen in the individual infrared images. The appearance of this spiral-ring varies remarkably from one band to another (Forbes et al. 1992). Whereas the appearance in the optical bands is dominated by emission line regions and dust extinction, the infrared light comes almost exclusively from stars and is affected to a much lesser extent by absorption than the optical light (Frogel 1985, 1988). Table 2 (see column labeled 'ring') indicates that the $JHK$ colors of this ring are several tenths of a magnitude redder than any other part of the galaxy except for the nucleus. The optical colors vary by as much as two magnitudes in the ring (see Figure 4). This range extends from some of the bluest to the reddest colors observed in the galaxy. These large color variations probably result from both high extinction and a population of young stars (Forbes et al. 1992).

Outside the ring and dust lanes (labeled under column 'bulge' in Table 2) the optical colors are somewhat redder than further out in the bar (labeled under column 'bar' in Table 2). $J - K$ also is slightly redder in the bulge than further out in the bar. These color changes are either from a redder bulge population or because there is a large amount of evenly distributed dust in the the bulge region.

We may use the $K$ brightness of the inner spiral to estimate the mass of the stars in it. We estimate that the difference in surface brightness on and off the spiral structure corresponds to $\sim 16.4$ mag/arcsec$^2$ in the ring. If we assume that the ring has a width of $\sim 4'' = 320$pc and a radius of $\sim 10'' = 800$pc then we estimate a total $K$ magnitude of 10.5 for the ring or an absolute magnitude of $-20.5$. To estimate the mass in the ring we use mass-to-light ratios for a range of stellar populations listed in Worthey (1994). For a young stellar population 1.5 Gyr old of solar abundance which is predicted to have a mass-to-light ratio at $K$ of $0.4 L_K/M_\odot$ (Worthey 1994, for $L_K$ in solar K luminosities), we estimate a mass of $\sim 10^9 M_\odot$ in the ring. An older stellar population would have a mass-to-light ratio that is twice as large. The red colors and high gas content in this ring imply that there is probably significant extinction of the light, so that the true mass should be even higher than we have estimated. For comparison, the molecular gas mass in the ring is $\sim 1.3 \times 10^9 M_\odot$ (Gerin et al. 1988). Thus we find that the mass in the ring in stars is of the same order as that in gas. Because of the fine structure we see in the spiral (and we are limited by the resolution of our images) the spiral must have a far lower scale height ($\lesssim 3'' \sim 240$pc) compared to nearby disk or bulge stars.

### 3.4 The Inner Bar

The near infrared images reveal the presence of a small bar inside the spiral ring of length $\sim 20'' \sim 1.6$kpc approximately equal to the diameter of the ring. Nuclear bars and rings are expected to occur near the location of the ILR (Inner Lindblad Resonance) of a major bar (e. g. Kormendy 1982). This inner bar has a major axis position angle (from north) that is $28° \pm 3$, and an axis ratio that is less than 0.6. For comparison, the position angle of the major axis of the main outer bar is $148° \pm 3$. For a representation of the position angles of these bars see Figure 5. Knowing the inclination of the galaxy ($\sim 45°$) and position angle of its major axis at large radii ($\sim 134°$, Ondrechen and van der Hulst, & Hummel 1989), we may estimate the angles of the major axes of the two bars in the principal plane of the galaxy. In a coordinate system with angle zero along the major axis of the galaxy, the angles of the major axes of the inner and outer bars are $74°$ and $16°$ on the sky respectively, which project to angles of $78°$ and $20°$ in the principal plane of the galaxy. In this plane, the major axes of the two bars are therefore $\sim 60°$ apart and are not perpendicular. Because this inner bar is probably triaxial, projection effects should cause the major axis position angle we measure from the $K$ image to be somewhat different than its projected principal major axis or axis of symmetry. In other words, it is possible that the two bars could be perpendicular but that projection effects have affected our measurements of their major axes. However, using the formulae from deZeeuw & Franx (1989) we find that it is impossible to achieve the observed angle differences given the orientation of the galaxy and the bar axis ratios. We therefore conclude that this inner bar is indeed not perpendicular to the outer bar.

Stellar orbits within an ILR are expected to be perpendicular to the main outer bar. One possible explanation for the offset of the inner bar with respect to the main outer bar may be that there is a torque between the inner bar and the spiral arms in the star forming ring. This situation would be consistent with the simulations of Shaw et al. (1993) who found that isophote twists in the nuclear region of NGC 1097 as well as other barred galaxies may be due to torques between the gas component and the stellar component. A self gravitating inner bar may also have a pattern speed unrelated to that of the outer bar (Pfenniger & Norman 1990).

### 3.5 The Dust Lanes

The dust lanes are clearly seen in optical color maps, but are barely visible in the $J - K$ color map (see Figures 3 and 4) on both the south-eastern and the north-western side of the bar. On the western side, the dust lane extends into the ring (previously noted in optical images by Hummel, van der Hulst & Keel 1987). These dust lanes are remarkably free of star formation except for small spots of H$\alpha$ emission towards the end of the bar that are coincident with a few blue spots in the dust lanes (see the H$\alpha$ image displayed in Figure 4; the H$\alpha$ image was provided by W. C. Keel and is described in Hummel, van der Hulst & Keel (1987)). Slices oriented perpendicular to the bar in all bands are shown in Figure 6. We find that the depth of the dust lanes in NGC 1097 decreases as a function of wavelength, is largest at $B$ and smallest at $H$ and $K$. This is expected for colors determined purely by dust extinction at low total opacity. However, this is not what was observed in the spiral arms of M51 (Rix & Rieke 1993), where the spiral arm-interarm contrast was minimized at $I$ band, and was larger at both bluer ($V$ and $B$ bands) and redder wavelengths ($J$,$H$ and $K$ bands). It may be that the higher dust opacity in M51 and the large effect of scattering at high opacity caused the color differences in M51; alternatively changes in stellar population as a function of both depth and position in the galaxy could be responsible.

We may put a limit on the maximum opacity of the dust lanes based upon their near infrared colors. We measure a maximum difference on and off the dust lanes in $J - K$ of 0.1 which is equivalent to a maximum extinction $A_V \sim 0.35$ using the galactic reddening law of Mathis (1990). If the dust is mixed with stars, or if there is a uniform component of dust outside the dust lanes, then $A_V = 0.35$ is an underestimate of the maximum total dust opacity . However, the sharpness of the dust lanes implies that the dust has a low vertical scale height. Typically dust has a vertical scale height that is smaller than the stellar scale height, so that we can think of the dust as a thin layer located between two thicker layers of stars. For a dust lane located on the near side of the galaxy, the dust will act like a screen, and our estimate of the maximum total dust opacity is correct (see below for a comparison of the two dust lanes). We cannot determine from the color maps whether there is a uniform distribution of dust outside the dust lanes. We note, however, that fluid simulations (e. g. Athanassoula 1992, van Albada & Roberts 1981) predict that dust lanes should have a far higher gas surface density than regions outside them. This suggests that there is little dust outside the dust lanes so that our limit $A_V < 0.35$ in the dust lanes is correct. The low opacity of the dust lanes is consistent with the fact that they are difficult to see in the near infrared images (see Figure 6).

We note that the dust lanes on the north western side of the galaxy are more prominent in the color maps than on the south eastern side. The position angle of the major axis of the galaxy is almost parallel to the bar, so that it is likely that the south-western side of the galaxy is nearer to us than the north-eastern side. Color-color plots are shown in Figure 7 for four strips free of foreground stars that lie across the bar at a position angle perpendicular to the major axis of the galaxy. The dust lanes on the near side of the galaxy (seen on the north western side of the bar) appear as straight lines in the color color plots. The slopes of the points in these plots are the same as those given by galactic extinction values (Mathis 1990) so that the colors are consistent with a dust screen in front of star light. This is consistent with the hypothesis that this dust lane is on the near side of the galaxy. However the color-color plots on the other side of the galaxy appear to have shallower slopes which would indicate that we are seeing a stellar component of the galaxy in front of the dust (e. g. Witt et al. 1992), consistent with the hypothesis that this dust lane is on the far side of the galaxy. We note that from Figure 7, it is difficult to determine the background colors of the galaxy. In fact, outside the dust lanes there are significant optical color gradients both along the bar, and perpendicular to the bar. Because of the difficulty of separating changes in the underlying stellar population from the effects of dust absorption and scattering, we did not find it possible to accurately determine the column depth of dust as a function of position in the galaxy.

We note that we do resolve the dust lanes in the $B - J$ and $B - K$ color maps; in other words they are wider than $4 - 8'' = 320 - 640$pc. They reach their peak depth at a position that is not coincident with with spots of H$\alpha$ emission (see Figure 4). Previous studies of spiral galaxies find that in their outer parts the H$\alpha$ emission is usually associated with dense molecular gas (Jean Turner private communication, e. g. Quillen et al. 1994). The presence of some blue stars might have reduced the extinction at the location of the H$\alpha$ spots, or alternatively the dense gas could be confined to an extremely narrow region so that higher resolution images would be required to detect the peak extinction in the color maps.

### 3.5.1 Comparison to radio continuum

Ondrechen and van der Hulst (1983) noted that the radio continuum emission at 1.4 GHz was coincident with the dust lanes. Here we compare the hydrogen column depth predicted from the dust extinction to the emissivity of the radio continuum maps. Because of the low opacity and level of star formation in the dust lanes, the colors of the dust lane on the north west (near) side of the galaxy, because it behaves similar to a dust screen, should be a reasonable tracer of hydrogen column depth. In Figure 8 we show a cross section along a position angle 44° (perpendicular to the major axis of the galaxy at large radii) $18'' = 1440$pc from the nucleus where the double peaked nature of the dust lane is apparent. The peaks are about $5''$ or 400pc apart. The dip has extinction in magnitudes that is $\sim 90\%$ of the peak extinction. The radio continuum at 1.5 GHz at this position is also double peaked with the dip perhaps $\sim 85\%$ of maximum intensity (from Figure 9 of Ondrechen van der Hulst and Keel 1987). The two ratios are remarkably close. (We note, however, that outside the dust lane the radio continuum emissivity and the dust opacity are not known, so that it could be a coincidence that the two ratios are remarkably close.)

That the radio continuum and the dust features are correlated (as has been observed in a number of galaxies, e. g. Ondrechen 1985) suggests that the cold atomic component and the cosmic ray component of the ISM are moving together so that the column density in these two phases are proportional. This is one of the assumptions of the shock models studied by Roberts & Yuan (1970). However, the non-thermal radio continuum from synchrotron emission is sensitive to both the density of cosmic ray electrons and the strength of the magnetic field, so the near proportionality of extinction to radio continuum emissivity is unexpected. Since there is no evidence of star formation in this region (see beginning of §3.5), it is unlikely that this near proportionality is caused by the effects of local star formation.

For cosmic ray electrons with a power law energy spectrum, the synchrotron emissivity at frequency $\nu$ is $\propto \rho_e B_\perp^{1-\alpha} \nu^\alpha$ for $B_\perp$ the line of sight component of the magnetic field, $\alpha$ the spectral index, and cosmic ray electron density $\rho_e$. Using Hummel, van der Hulst & Keel (1987)'s estimate for the spectral index $\alpha \sim -0.77$ we find that emissivity $\propto \rho_e B_\perp^{1.77}$. Note that the emissivity is is reasonably sensitive to $B$. Because the position angle of the major axis of NGC 1097 is close to the major axis of the bar, we are not sensitive to the uniform magnetic field component that is parallel to the shock. However, the emissivity should be sensitive to the uniform magnetic field component perpendicular to the shock as well as the non-uniform component, which is typically observed to be about half of the uniform component (e.g. Goodman & Heiles 1994). Simple shock models (e. g. Roberts & Yuan 1970, Ondrechen 1985) predict that $B \propto \rho_e$ with $B$ close to parallel to the shock. The near proportionality of the non-thermal radio continuum emissivity and the dust extinction puts a limit on the change in amplitude of the uniform magnetic field component perpendicular to the shock as well as the non-uniform component. If the cosmic ray electron density is proportional to the hydrogen density then the near proportionality of the non-thermal radio continuum with the dust extinction suggests that these two components of the magnetic field do not change across the shock. However, then the cosmic ray pressure would not remain in equipartition with the magnetic pressure on both sides of the shock (unless the non-uniform component of the magnetic field is small). Alternatively if both the cosmic ray electron density and magnetic field change in the shock so that equipartition is maintained, then the cosmic ray electron density can not be proportional to the hydrogen density and we conclude that the two phases of the ISM do not move together. We note that Tilanus & Allen (1988) find that although the non-thermal component of the radio continuum in M51 peaks at the location of the dust lanes, the slope is shallower in the preshock side of the spiral arm than the postshock side. They find this inconsistent with density wave models and suggest that the cosmic ray component has decoupled from the cold gas component. A more detailed study of the radio continuum in the dust lanes in NGC 1097 or another similar galaxy with little star formation in the dust lanes (making it easier to both estimate extinction and the non-thermal component of the radio continuum) could resolve the role of the magnetic field in these shocks, as well as determine whether the cosmic rays move together with the colder gas traced in extinction from dust.

## 4. CONCLUSION AND SUMMARY

In this paper we have presented images of NGC 1097 in $B, V, R, J, H$ and $K$. The inner spiral has large color variations from both extinction and young stars. Inside this inner spiral ring we find an inner bar that is *not* perpendicular to the main outer bar in the principal plane of the galaxy, as would be expected for stellar orbits inside an ILR. This suggests that there are torques between the inner bar and the spiral ring (e. g. Shaw et al. 1993) and that this inner bar may have a pattern speed different than the outer bar (Pfenniger & Norman 1990). We estimate that the stellar mass of the star forming ring is of the same order of magnitude as the mass of the molecular gas. Future high angular resolution studies of the spiral ring and inner bar may be able to determine the degree of interaction between these two components.

The dust lane on the north-west side of the galaxy has optical and infrared colors of a dust screen with opacity derived with a galactic extinction law (Mathis 1990). This is consistent with this dust lane being located on the near side of the galaxy. Colors in the opposite dust lane are consistent with dust sandwiched between stars, suggesting that this dust lane is located on the far side of the galaxy. Although these dust lanes are free from star formation, because of the difficulty of separating changes in the underlying stellar population from the effects of dust absorption and scattering, we did not find it possible to accurately determine the column depth of dust as a function of position in the galaxy. However based upon their near infrared colors, we find that they are low opacity with $A_V < 0.35$.

Near the inner spiral (at $\sim 17'' = 1360\text{pc}$ from the nucleus) the dust lane is double peaked with peak to valley extinction ratio similar to that observed in radio continuum by Ondrechen and van der Hulst (1987). This suggests that if the cold diffuse component of the ISM (traced by the dust) moves with the cosmic ray electron component then the magnetic field pressure is not a significant force in the shocks. Alternatively if the two phases of the ISM do not move together then a more detailed comparison should show that the shapes of the shocks observed in these two phases should differ. Recent improvements in the sensitivity of radio observations will make it interesting to do a careful comparison of dust extinction maps with radio continuum maps at different wavelengths. In particular dust features such as those in NGC 1097 with little star formation could be studied and compared to studies of the spiral arms in galaxies such as M51 (Tilanus & Allen 1988)

We thank W. C. Keel for providing us with the H$\alpha$ image of NGC 1097. We acknowledge helpful discussions and correspondence with R. White, R. W. Pogge, K. Sellgren, and A. Gould. J. A. F. thanks the staff of the Las Campanas Observatory in Chile for their assistance and Leonard Searle for providing access to the research facilities of the Carnegie Observatories. The OSU galaxy survey is being supported in part by NSF grant AST-9217716. The Las Campanas Observatory IR camera was built with partial funding from NSF grant AST-9008937 to S. E. Persson. J.A.F's research is supported in part by NSF grant AST-9218281. We thank the anonymous referee for helpful comments and suggestions which improved the paper. LEK's and DT's observing in Chile was supported by NSF grant AST91-57038 to DT. A.C.Q. acknowledges the support of a Columbus fellowship.

TABLE 1
Summary of Las Campanas Infrared Observations

| Band | Position | Date | Total Exptime |
|---|---|---|---|
| $J$ | Nuc | 1992 Nov. 4 | 21 minutes |
| $K$ | Nuc | 1992 Oct. 31 | 18 minutes |
| $J$ | Nuc | 1993 Oct. 3 | 2.5 minutes |
| $H$ | Nuc | 1993 Oct. 3 | 2.5 minutes |
| $K$ | Nuc | 1993 Oct. 3 | 2.0 minutes |
| $J$ | NE | 1992 Nov. 4 | 20 minutes |
| $K$ | NE | 1992 Nov. 7 | 21 minutes |
| $J$ | NW | 1992 Nov. 4 | 18 minutes |
| $H$ | NW | 1993 Oct. 3 | 18 minutes |
| $K$ | NW | 1992 Nov. 4 | 15 minutes |
| $K$ | NW | 1992 Nov. 7 | 27 minutes |
| $J$ | SE | 1992 Nov. 4 | 20 minutes |
| $K$ | SE | 1992 Nov. 7 | 45 minutes |
| $J$ | SW | 1992 Nov. 4 | 20 minutes |
| $K$ | SW | 1992 Nov. 7 | 45 minutes |

TABLE 2
Colors [1]

| color | nucleus | bulge | ring[2] | bar | NGC 1097A |
|---|---|---|---|---|---|
| $B - K$ | 5.36 | 4.49 | 3.2-5.0 | 4.12 | 3.7 |
| $V - K$ | 4.20 | 3.44 | 2.9-4.0 | 3.20 | 2.76 |
| $R - K$ | 3.49 | 2.82 | 2.5-3.3 | 2.67 | 2.22 |
| $J - K$ | 1.24 | 0.87 | 1.0-1.2 | 0.85 | 0.77 |
| $H - K$ | 0.56 | 0.18 | 0.25-0.50 | 0.17 | 0.18 |

[1] Estimated errors are typically 0.1 mag. [2] Range of colors observed in the star-forming spiral ring. In this ring there are very blue spots as well as very red regions.

# FIGURE CAPTIONS

**Figure 1.** Images of NGC 1097 at $B$, $V$, $R$, $J$, $H$, and $K$ displayed in mag binned to a pixel scale of $1.8''$. North is up, east is to the left and the scale in arcsecs is shown on the lower left. Contours are shown at 0.5 mag intervals with highest contours at 23.5, 23.0, 22.5, 21.5, 20.0, 19.5 mag arcsec$^{-2}$ at $B,V,R,J,H$, and $K$ respectively. NGC 1097A is the dwarf galaxy to the north west of NGC 1097. All images have been transformed to the same pixel scale, orientation and resolution. Note that in the optical bands both the effects of dust absorption and young blue stars can be seen. The infrared images have far less small scale structure.

**Figure 2.** Images in magnitudes of NGC 1097A, the dwarf elliptical galaxy north west of NGC 1097. Contours are shown at 0.5 mag intervals with lowest contour 19.5 mag arcsec$^{-2}$ at $B$ and 15.5 mag arcsec$^{-2}$ at $K$. Also shown are $B-K$ and $J-K$ color maps with a range displayed from 0.2 to 1.2 mag at $J-K$ and 3 to 4 mag at $B-K$. Note that the colors maps are remarkably smooth. The scale in arcsecs is shown on the lower left.

**Figure 3.** Images of the nuclear region of NGC 1097 at $B$, $V$, $R$, $J$, $H$, and $K$ linear in intensity. Also shown to the same scale is an H$\alpha$ +[NII] image and $J-K$ and $H-K$ color maps in magnitudes. The H$\alpha$ image was provided by W. C. Keel and is described in Hummel, van der Hulst & Keel (1987). The total H$\alpha$ luminosity is $3.4\ 10^{-12}$ erg cm$^{-2}$ s$^{-1}$, and a range of 0 to $1.9\ 10^{-14}$ erg cm$^{-2}$ s$^{-1}$ arcsec$^{-2}$ is displayed. The scale in arcsecs is shown on the lower left.

**Figure 4.** Color maps of NGC 1097 displayed in magnitudes. Gray scale range is displayed from 3.0, 1.8, 0.3, -0.2, 0.2 mag with a range of 1.9, 1.7, 0.77, 0.61, 1.13 mag at $B-K$, $R-K$, $J-K$, $H-K$ and $B-V$ respectively. An H$\alpha$ + [NII] image is included on the top right corner, which was provided by W. C. Keel and is described in Hummel, van der Hulst & Keel (1987). The scale in arcsecs is shown on the lower left.

**Figure 5.** Position angles of the outer and inner bar on the sky. North is up and east is to the left. The bar of NGC 1097 has major axis with a position angle (clockwise from north) of $148°$ and is represented by the large ellipse. The star forming ring is represented by the circle. The inner bar is inside the star forming ring and has major axis with a position angle of $28°$. The major axis of the galaxy at large radii has a position angle of $134°$ (Ondrechen et al. 1989) and is displayed as a dotted line though the nucleus.

**Figure 6.** Slices across the bar. In order of height the lines in each plot are $K$ (top), $J$, $R$, $V$, and $B$ (bottom). The arrows in each plot show the location of the dust lane. These plots are made from averages of intensity in strips that are $5''$ wide oriented along a position angle (clockwise from north) of $44°$ which is the position angle of the minor axis of the galaxy at large radii (Ondrechen, van der Hulst & Hummel 1989), and is approximately perpendicular to the bar (see Figure 5). The horizontal axis is in arcsecs, and the vertical axis is in mag arcsec$^{-2}$. The distance of each strip from the nucleus is shown in the top left hand corner of each plot. The two plots on the left side are from the north west side of the galaxy and the plots on the right side are on the opposite side of the galaxy. The two plots on the left are labeled 'near side' since the dust lanes are probably nearer to us than the dust lanes on the opposite side.

**Figure 7.** Color-color plots for strips perpendicular to the bar, showing cross sections of the dust lanes. The points are from strips that are $5''$ wide oriented along a position angle (clockwise from north) of $44°$ from north which is the position angle of the minor axis of the galaxy at large radii

(Ondrechen, van der Hulst & Hummel 1989), and is approximately perpendicular to the bar. Axes are in mag and the distance of each strip from the nucleus is shown to the right hand side of each plot. For a dust screen in front of starlight the points should lie along a straight line. The slope for a dust screen with a typical galactic reddening law (Mathis 1990) is shown by the lines in the lower right hand corner of each plot. For dust mixed with stars the slope of the line will be shallower. The color-color plots from the north-western lane (shown on the left hand side) have a slope consistent with a dust screen. This suggests that this dust lane is nearer to us than the opposite dust lane.

**Figure 8.** Cross section of the double peaked portion of the dust lane. The horizontal axis is in arcsecs, and the vertical axis is in $B - K$ mag. This plot is from a slice perpendicular to the dust lane $18''$ from the nucleus. The slice is $\sim 2''$ wide and is on the north-west side of the galaxy oriented along a position angle (clockwise from north) of $44°$ which is the position angle of the minor axis of the galaxy at large radii (Ondrechen, van der Hulst & Hummel 1989), and is approximately perpendicular to the bar (see Figure 5).